# Kinetics of the simplest Criegee intermediate reaction with ozone studied by mid-infrared quantum cascade laser spectrometer

Yuan-Pin Chang[a,b,*], Hsun-Hui Chang[a], and Jim Jr-Min Lin[a,c]

[a] *Institute of Atomic and Molecular Sciences, Academia Sinica, Taipei 10617, Taiwan*

[b] *Department of Chemistry, National Sun Yat-sen University, Kaohsiung 80424, Taiwan*

[c] *Department of Chemistry, National Taiwan University, Taipei 10617, Taiwan*

AUTHOR INFORMATION

**Corresponding Author**

* To whom correspondence may be addressed. Email: ypchang@mail.nsysu.edu.tw

ABSTRACT

The kinetics of $CH_2OO$ reaction with ozone has been studied by monitoring $CH_2OO$ with time-resolved infrared (IR) absorption spectroscopy, which utilized fast chirped IR pulse train from a quantum cascade laser [J. Chem. Phys., 2017, **146**, 244302]. $CH_2OO$ was prepared by photolyzing a gas mixture of $CH_2I_2/O_2/O_3$ at 352 nm; the photolysis wavelength was chosen to minimize the photodissociation of $O_3$. The measured rate coefficient at 298 K and 30 Torr is $(6.7\pm0.5)\times10^{-14}$ cm$^3$sec$^{-1}$, independent of pressure from 30 to 100 Torr. The result indicates that previous *ab initio* calculations either underestimated or overestimated this reaction rate by one order of magnitude or more. The result also implies that in laboratory studies of ozonolysis of alkenes, the reaction of Criegee intermediate with ozone may play a role. However, this reaction would not compete with other $CH_2OO$ sinks in the atmosphere.

**Introduction**

In the atmosphere, ozonolysis of alkenes produces highly reactive Criegee intermediates (CIs).[1–4] The produced CIs may have excess internal energy and undergo unimolecular processes, such as isomerization or decomposition to form OH radicals; some of the CIs may be collisionally stabilized and then may react with other atmospheric species. In laboratory studies of ozonolysis, direct detection of CIs is difficult due to low steady-state concentrations. Direct spectroscopic and kinetic studies of CIs only became feasible after the work by Welz *et al.*,[5] which demonstrated the efficient preparation of the simplest CI, $CH_2OO$, via the reaction of $CH_2I$ + $O_2$ → $CH_2OO$ + I. Many recent studies on prototypical CIs have found that CIs play an important role in atmospheric chemistry, including formation of OH radicals[6–9] and oxidation of atmospheric species, like $SO_2$, $NO_2$, organic and inorganic acids, alkenes, and water vapor.[5,10–31] The reactions of CIs may produce radicals or low-volatility organic species, which are key components in the formation of secondary organic aerosols.[32]

The reactions of $O_3$ with CIs are potentially important in both laboratory and atmospheric studies. For example, Novelli et al.[7] have considered that the main loss paths of CIs in their ozonolysis experiments are unimolecular decomposition and reaction with ozone at early reaction times, whereas reaction with organic peroxy radicals, alcohols, aldehydes and organic peroxides become more important at later reaction times. However, to the best of our knowledge, there is no direct measurement on the rates of ozone reaction with CIs, while there have been a few theoretical works.[33–36] For $CH_2OO$, Kjaergaard et al.[33] calculated its reaction with $O_3$ by using CCSD(T)//B3LYP level of theory and predicted the formation of a cycloaddition intermediate through a significant barrier, yielding small rate coefficients ranging from $4.0\times10^{-16}$ to $1.2 \times 10^{-18}$ cm$^3$ sec$^{-1}$. Wei et al.[34] also investigated the reaction mechanism by using CCSD(T)//B3LYP level of theory, but they did not report any rate coefficient. Vereecken et al.[35,36] calculated this reaction by using CCSD(T)//M06-2X level of theory, predicting that CH$_2$OO and O$_3$ will firstly form a prereactive complex without a barrier before passing a submerged chain-addition transition state; they predicted a larger rate coefficient of $4 \times 10^{-13}$ cm$^3$ sec$^{-1}$. It appears that it is not easy to accurately estimate the rate coefficient with present quantum chemistry approaches. Nonetheless, all the theoretical studies predicted CH$_2$O + 2O$_2$ as the final products.[33–36]

Recently, we have developed a high-resolution mid-infrared quantum cascade laser (QCL) spectrometer, and utilized it to study the spectrum of the $\nu_4$ band of CH$_2$OO.[37] In this work, we used the QCL spectrometer to study the kinetics of CH$_2$OO reaction with O$_3$. This method has the following advantages: (1) avoiding byproduct interferences and baseline drifting by probing narrow spectral lines, (2) high sensitivity due to narrow laser linewidth and long optical path length, (3) long CH$_2$OO lifetimes (up to 14 ms) due to low required concentration of CH$_2$OO, and (4) efficient data acquisition using fast chirped pulse train of QCL.

**Experimental methods**

Most of the experimental procedures have been described in our previous work.[37] Thus only a brief description is provided here. CH$_2$OO was prepared in a flow cell following the well-established method of CH$_2$I$_2$/O$_2$ photolysis:[5] CH$_2$I$_2$ (1–7 mTorr) mixed with O$_2$ (30 or 100 Torr) was photolyzed by an unfocused excimer laser beam at 352 nm (laser fluence: $(1.3–4.4) \times 10^{16}$ photon cm$^{-2}$). We chose this photolysis wavelength (instead of 248 or 308 nm) to minimize the effect of $O_3$ photolysis (cross section of $O_3 \approx 1.0 \times 10^{-22}$ cm$^2$ at 352 nm, probability of photolysis $\leq$ $4.4 \times 10^{-6}$ at $\leq 4.4 \times 10^{16}$ photon cm$^{-2}$). The $O_3$ gas was synthesized by using a commercial ozone generator, collected by adsorption on silica gel at the dry-ice temperature, and further purified by condensation in a stainless steel cylinder at the liquid-nitrogen temperature, following the procedures described in our previous work.[38] The $O_3$ concentration was measured via its UV absorption by using an absorption cell with a path length of 5 cm, a Deuterium lamp (Hamamatsu, L10671D), and a spectrometer (Ocean Optics, USB2000+UV-VIS-ES), right before the $O_3$ gas entering the reactor cell. The flow rates of the used gases were controlled by mass flow controllers (Brooks, 5850E).

We used a distributed-feedback quantum cascade laser (Alpes Lasers, CW-DFB-QCL) as a coherent IR source to probe $CH_2OO$. The laser was driven by an intermittent CW driver (Alpes Lasers), powered by a DC-power supply (HAMEG Instruments, HMP2020). Its frequency coverage is 1279–1290 cm$^{-1}$ and the practical spectral linewidth is about 0.002 cm$^{-1}$ with a peak power over 10 mW. The temperature of the laser was controlled by a TEC cooling element inside the laser housing and a temperature controller. The laser was operated in pulse-mode with a period of 18 $\mu$s or 120 $\mu$s and 40% duty cycle (7 $\mu$s or 48 $\mu$s pulse duration). In each laser pulse, the laser frequency was down-chirped (1286.1–1285.5 cm$^{-1}$ in 48 $\mu$s). During the experiment, the IR chirped pulses were repeatedly scanned through the Q branch of the $\nu_4$ fundamental band as shown in Figure 1. The lifetime of $CH_2OO$ in our experiments was in the millisecond time scale, much longer than the IR pulse period. Thus, for a single photolysis event, we could obtain a full time profile of $CH_2OO$ from probing its transitions with the IR pulse train.

Our reactor cell was a glass tube (inner diameter 19 mm) with $BaF_2$ windows at both ends, which were purged with $N_2$. The photolysis excimer laser beam (352nm, repetition rate 1 Hz) was combined with the IR probe beam and introduced into the flow cell by a high reflective $BaF_2$ mirror at 352 nm (Eksma Optics, custom item). A $BaF_2$ right-angle prism and a concave aluminium mirror (Edmund Optics, part # 43549) were placed before and after the flow cell, such that the probe IR beam was reflected back-and-forth between the two optics, creating a long overlap with the UV photolysis volume up to 3.9 meters (6 passes though the cell with an effective sample length of 65 cm). After leaving the multi-pass cell, the IR probe beam was guided to a HgCdTe (MCT) detector (Kolmar Technologies, KMPV11-1-J2), and the UV photolysis laser beam was reflected away by another $BaF_2$ high reflective mirror. Each photolysis pulse was synchronized with the rising edge of one of the IR probe pulses by using a delay generator (SRS, DG535). The calibration of the IR laser wavelength was carried out by measuring a reference gas spectrum (3 Torr $N_2O$) and an etalon signal (Ge etalon 3" in length, Free Spectral Range = 0.0163 $cm^{-1}$). The IR pulse train signals from the DC outputs of all MCT detectors were acquired by an oscilloscope (LeCroy, HDO4034, 12-bit vertical resolution) at a sampling rate of 1.25GS/s. For each [$CH_2OO$] time profile measurement, we averaged the data for 100 UV laser shots to improve the signal-to-noise ratio.

Finally, we also utilized a UV transient absorption spectrometer to roughly estimate the absolute values of [$CH_2OO$] under the experimental conditions of this work. A LED (Hamamatsu, LC-L2) with an emission profile centered at 365 nm was used as the probe light source. The probe light was projected into the flow cell by a convex lens, and then it was reflected once by the concave mirror to achieve two passes through the cell. After leaving the flow cell, the remained probe light was guided to a balanced photodiode detector (Thorlab, PDB450A). The rest of experimental configurations were the same as those described above. Since the system was not optimized for the UV detection, the uncertainty in the absolute values of [$CH_2OO$] would be a bit larger than those of our previous works.[12]

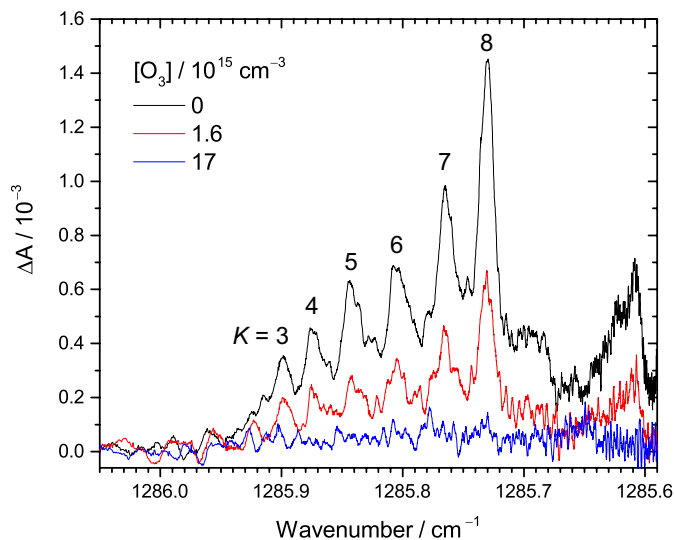

**Figure 1.** Examples of transient IR absorption spectra of $CH_2OO$ at different $O_3$ concentrations at 30 Torr total pressure. The peaks correspond to the Q branch of the $\nu_4$ fundamental band of $CH_2OO$. The $K$ number is assigned to each (partially) resolved sub-band. The photolysis-probe delay time is 2.4 ms in this example. See Table S1 (expt. 1) for details of the experimental conditions.

**Results and discussion**

Figure 1 shows selected transient IR absorption spectra of $CH_2OO$ at representative $O_3$ concentrations. The transient absorption means the change in IR absorption intensity with respect to that before the UV photolysis. In this example, the results at a photolysis-probe delay time of 2.4 ms are shown. As mentioned in Experimental methods, the spectral range corresponds to the Q branch of the $\nu_4$ fundamental band of $CH_2OO$.[37] Note that each of the peaks between 1285.9 and 1285.7 cm$^{-1}$ corresponds to congested transitions from the same *K* levels (from *K* = 3 to 8), while the broad feature peaked at 1285.62 cm$^{-1}$ is the perturbed transitions of higher *K* levels.[37] See Ref. 37 for detailed spectroscopic assignments and perturbation analysis. When increasing the $O_3$ concentration, the $CH_2OO$ signal decreases accordingly, suggesting $CH_2OO$ is consumed by its reaction with $O_3$. The decay trends of these peaks are similar, indicating no rotational dependence.

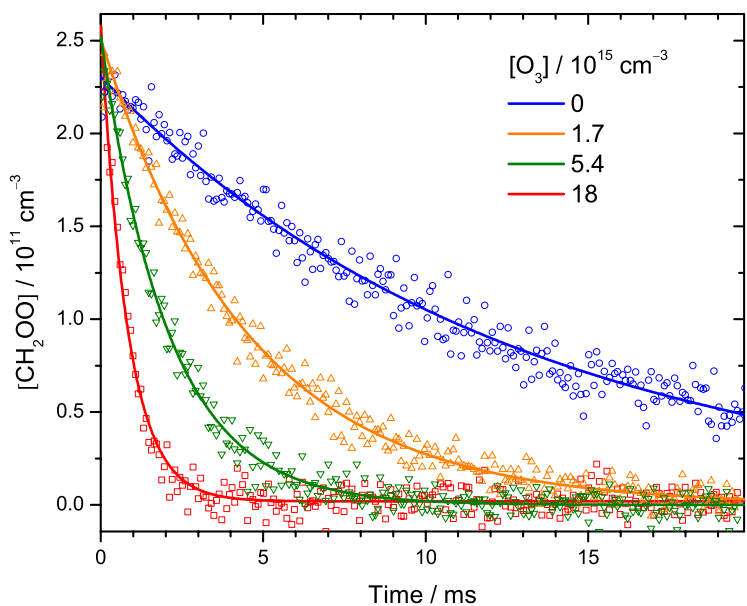

**Figure 2.** Representative time profiles of [$CH_2OO$] at various [$O_3$] at 30 Torr and 298 K. Each data point (symbol) is the integrated intensity of the highest peak (1285.71–1285.74 cm$^{-1}$) of the Q branch of the $CH_2OO$ $\nu_4$ band (see Figure 1) probed by each IR pulse. The lines are single-exponential fit to the data. See Table S1 (expt. 11) for details of the experimental conditions.

Figure 2 shows a few time profiles of [CH$_2$OO] at various [O$_3$]. The increasing decay rates at higher [O$_3$] indicate CH$_2$OO reacts with O$_3$. The following reactions are involved in the formation and decay of CH$_2$OO in our CH$_2$I$_2$/O$_2$/O$_3$ photolysis system:[12]

CH$_2$I$_2$ + $h\nu$ (352 nm) → CH$_2$I + I

CH$_2$I + O$_2$ → CH$_2$OO + I

CH$_2$OO → products $\qquad\qquad\qquad$ $k_{first}$

CH$_2$OO + O$_3$ → products $\qquad\qquad$ $k_{O3}$

CH$_2$OO + I → products $\qquad\qquad$ $k_I$

CH$_2$OO + CH$_2$OO → products $\qquad$ $k_{self}$

The pulse width of the photolysis laser was only 20 ns. Under our experimental conditions, very high O$_2$ concentrations (30 Torr or more) were used, leading to fast conversion of CH$_2$I to CH$_2$OO (~1 $\mu$s). Thus, we neglect the time of CH$_2$OO formation in our kinetic analysis.

In our first attempt of data analysis, we described the measured time profile of CH$_2$OO as:

$$[CH_2OO](t) = [CH_2OO]_0 \exp(-k_{eff}t) \qquad\qquad\qquad (1)$$

$$-\frac{d[CH_2OO]}{dt} = k_{eff}[CH_2OO] = (k_0 + k_{O3}[O_3])[CH_2OO] \qquad\qquad (2)$$

Where $k_{eff}$ is the observed 1st-order rate coefficient of CH$_2$OO decay, $k_{O3}$ is the bimolecular rate coefficient for the reaction of CH$_2$OO with O$_3$ and $k_0$ is the decay rate coefficient of CH$_2$OO without O$_3$, which may include the contributions from first-order loss $k_{first}$ (mostly wall loss), self-reaction ($k_{self}$), and reaction with other radicals (mostly iodine atoms, represented by $k_I$).

$$k_0 \simeq k_{first} + k_I[I] + 2k_{self}[CH_2OO] \qquad\qquad\qquad (3)$$

Solid lines in Figure 2 are the fits to the transient absorption time profiles with Equation (1), yielding $k_{eff}$ and [CH$_2$OO]$_0$. As mentioned in our previous work,[37] we also measured the UV absorbance of $CH_2OO$ at 365 nm (where the absolute UV cross section is known)[39] under similar experimental conditions, which allows us to roughly estimate the absolute value of [$CH_2OO$] in the IR experiments.

Figure 3 plots the value of $k_{eff}$ as a function of [$O_3$] at 30 Torr and 298 K, and the solid lines are linear fits to the data points (except those at [$O_3$]=0). At low [$CH_2OO$]$_0$ ($\approx 2.4 \times 10^{11}$ cm$^{-3}$, Figure 3(a)), $k_{eff}$ shows a linear relationship with [$O_3$], as expected from Equation (2). The slope ($6.50 \times 10^{-14}$ cm$^3$ sec$^{-1}$) should correspond to a measured value for $k_{O3}$. To our surprise, at higher [$CH_2OO$]$_0$ (for example, Figure 3(b)), while $k_{eff}$ still shows a linear function of [$O_3$] for [$O_3$]>0, this linear line does not go through the data points at [$O_3$]=0. Nonetheless, the $k_{eff}$ as a function of [$O_3$] all exhibit a consistent slope for [$O_3$]>0. The detailed values are listed in Table S1 in the column of $k_{O3}$. The averaged value (with one-sigma error bar) of the slope is $(6.72 \pm 0.46) \times 10^{-14}$ cm$^3$ sec$^{-1}$.

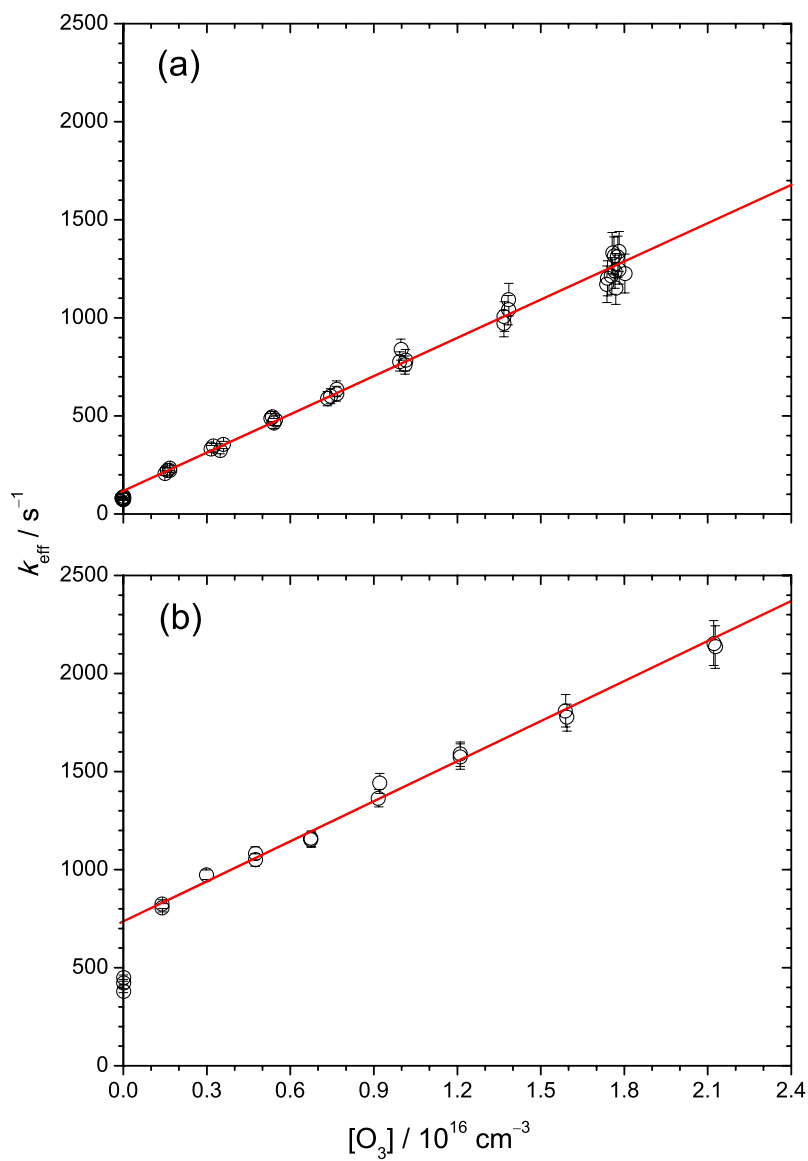

**Figure 3.** Effective first-order rate coefficient of $CH_2OO$ decay, $k_{eff}$, as a function of $[O_3]$ for $[CH_2OO]_0 =$ (a) $2.4 \times 10^{11}$ cm$^{-3}$ (expt. 11) and (b) $2.1 \times 10^{12}$ cm$^{-3}$ (expt. 9). The error bars are $1\sigma$ uncertainties from the fitting of the $[CH_2OO]$ time profiles. The red line is a linear fit to $k_{eff}$ excluding the data at $[O_3] = 0$ (see text for details). The slope of the red line corresponds to the second-order rate coefficient of $CH_2OO + O_3$ reaction.

To investigate the possible source that causes the difference between the intercept value ($k_{incpt}$) of $k_{eff}$ (obtained from the linear fit of $k_{eff}$ as a function of [$O_3$] for [$O_3$]>0) and the measured $k_0$ ($k_{eff}$ at [$O_3$]=0), we have performed several checking experiments as described below. First, one may wonder whether the O atoms from $O_3$ photodissociation affect the measurements or not. To check this, we measured $k_{eff}$ at different 352 nm laser fluences (by a factor of 2.8, which varied [O] by the same factor) and obtained very similar results (see Figure S1), indicating no $O_3$ photochemistry involved. As mentioned in Experimental methods, the probability of $O_3$ photolysis is less than $4.4 \times 10^{-6}$, thus the upper limit of [O] can be estimated to be $(4.4 \times 10^{-6})(2 \times 10^{16} \text{ cm}^{-3}) = 9 \times 10^{10} \text{ cm}^{-3}$. The rate of $CH_2OO$ reaction with O atom is not known yet. If we assume every collision between $CH_2OO$ and O atom leads to reaction (the collision-limit), the rate coefficient would be about $3 \times 10^{-10} \text{ cm}^3 \text{ sec}^{-1}$,[40] and the effect of the O atoms on the $CH_2OO$ decay is still small (< 27 sec$^{-1}$ in $k_{eff}$).

Because we synthesized and purified the $O_3$ gas by ourselves, one may worry about its purity. The typical lifetime of our $O_3$ gas (stored in a stainless steel cylinder at the dry-ice temperature) is more than 20 hours. To check the effect of the impurity, we deliberately warmed up the $O_3$ gas to room temperature and waited until all $O_3$ molecules had decomposed (~70 hours, verified with UV absorption). We called this gas "decomposed $O_3$" which would contain a similar or higher level of impurity, compared to that of our fresh $O_3$ gas. The results are shown in Figure S3. The "decomposed $O_3$" gas does not change $k_{eff}$ at all, indicating the impurity in our $O_3$ gas has negligible effect.

As shown in Figure 3, the difference between $k_{incpt}$ and $k_0$ is more significant at higher [$CH_2OO$]$_0$. We found this difference, $k_{incpt} - k_0$, is proportional to [$CH_2OO$]$_0$ as shown in Figure

S4. This observation gave us an idea that the chemistry may involve I atoms, of which the concentration is proportional to $[CH_2OO]_0$ in our preparation method. Thus, we propose the following reactions should also be involved when $O_3$ is present:

$O_3 + I \rightarrow IO + O_2$                                                       $k_{I+O3}$

$CH_2OO + IO \rightarrow$ products (possibly $CH_2O + OIO$)             $k_{IO}$

The reaction of I atom with $O_3$ has been well studied; the literature value of $k_{I+O3}$ is $(1.28\pm0.06)\times10^{-12}$ cm$^3$ sec$^{-1}$.[41] The lowest $[O_3]$ in our experiment is at least $1.6\times10^{15}$ cm$^{-3}$. Thus, I atoms would be quickly converted into IO due to the fast reaction rate ($k_{I+O3}[O_3] > 2000$ s$^{-1}$), which is much faster than the reaction of $CH_2OO$ with $O_3$ in our experiments. Since $[O_3] >> [I]$, the amount of IO is mainly controlled by the initial amount of I atoms, which is proportional to $[CH_2OO]_0$. Therefore, we may modify Equation (3) to Equation (4) for a better approximation of $k_{incpt}$.

$k_{incpt} \cong k_{first} + k_I[I] + 2k_{self}[CH_2OO] + k_{IO}[IO]$              (4)

And mass balance gives

$[I]_0 = [I] + [IO] \cong 2[CH_2OO]_0$                            (5)

If $k_{IO}$ is larger than $k_I$, we would have larger $k_{incpt}$ when $O_3$ is present.

We have also determined $k_{incpt}$ and $k_0$ at various $[CH_2OO]_0$ (see Table S1). Note that when $O_3$ is present, the maximum concentration of IO is also controlled by $[CH_2OO]_0$, $[IO]_{max} \cong [I]_0 \cong 2[CH_2OO]_0$. Therefore the difference between $k_{incpt}$ and $k_0$ would become:

$k_{incpt} - k_0 \cong (k_{IO} - k_I) [I]_0 \cong 2(k_{IO} - k_I) [CH_2OO]_0$.            (6)

As a result, the slope ($\sim 1.31\times10^{-10}$ cm$^3$ sec$^{-1}$) in Figure S4 provides an estimate for $2(k_{IO} - k_I)$. Assuming $k_I = 9.15\times10^{-11}$ cm$^3$ sec$^{-1}$,[12] the value of $k_{IO}$ is estimated to be on the order of $10^{-10}$ cm$^3$ sec$^{-1}$. Examples of kinetic simulation of the time profiles of the involved species can be found in the Supplementary Information (Figures S5 to S8). In the simulation, a value of $1.5\times10^{-10}$ cm$^3$ sec$^{-1}$ of $k_{IO}$ ($k_{10}$ in Supplementary Information, which uses different notations for the rate coefficients) gives satisfactory fits to the experimental data.

In brief summary, we have tested a few crucial experimental conditions. The results enable us to exclude the possibility of interferences which are originated from the photolysis of $O_3$ and the impurity in our $O_3$ gas. In our system, I atoms would be quickly converted into IO when $O_3$ is present; the side-reaction of $CH_2OO + I$ would be shifted to the reaction of $CH_2OO + IO$ and thus, changed the value of $k_{incpt}$. For $[O_3] > 1.6\times10^{15}$ cm$^{-3}$, all the observed data of $k_{eff}$ are linear with $[O_3]$ with a slope of $(6.72\pm0.46)\times10^{-14}$ cm$^3$ sec$^{-1}$, which corresponds to the bimolecular rate coefficient of $CH_2OO$ reaction with $O_3$. Finally, we found the kinetics at 100 Torr total pressure is very similar to that at 30 Torr (Figure S2), indicating weak pressure dependence in this pressure range.

As described in *Introduction*, the mechanism of this reaction and the rate coefficient have been predicted by Kjaergaard et al.[33] and Vereecken et al.[35,36] Our measured rate coefficient is more close to that from Vereecken et al., $4\times10^{-13}$ cm$^3$ sec$^{-1}$, which has the uncertainty of at least 1 order of magnitude.[35,36] While $O_3$ is isoelectronic with $CH_2OO$, the self-reaction rate of $CH_2OO$ ($\sim8\times10^{-11}$ cm$^3$sec$^{-1}$)[12] is at least 10 orders of magnitude larger than that of $O_3$. The reaction rate of $CH_2OO$ and $O_3$ is somehow in between these two limiting cases. While the zwitterionic character of $CH_2OO$ leads to a pure attractive cycloaddition in its self-reaction,[13,35] the interaction between $CH_2OO$ and $O_3$ would lead to the formation of a pre-reactive complex instead, which has a barrier for either subsequent chain addition or cycloaddition, according to the predictions by Vereecken et al.[35,36] A further clarification of the reaction mechanism via either theory or experiment is still required.

Finally, in our future work we would like to clarify the reaction mechanism via examining the reaction products. Theoretical works have predicted that $CH_2O$ and $O_2$ are the final products of this reaction. We plan to utilize our QCL spectrometer to probe $CH_2O$ created from the reactions, while a new QCL with a spectral range dedicated to detecting $CH_2O$ will be required. Vereecken et al.[35,36] further predicted that the oxygen atom in the product $CH_2O$ is purely from $O_3$, while the two oxygen atoms from $CH_2OO$ are released as $O_2$. However, the cycloaddition mechanism predicted by Kjaergaard et al.[33] would lead to a 1:1 ratio of oxygen atoms of $CH_2O$ originating from $CH_2OO$ or $O_3$.[36] To verify these predictions, our future works will also be isotopic labelling experiments and the IR identifications of isotope-substituted $CH_2O$.

**Conclusions**

In summary, we investigated the kinetic of $CH_2OO$ reaction with $O_3$ at 298 K and 30 Torr by using a high resolution mid-infrared quantum cascade laser spectrometer. No pressure dependence was observed from 30 to 100 Torr. Different from other reactant molecules like $H_2O$, $SO_2$, $NO_2$, and organic/inorganic acids, ozone would react with I atoms to form IO, an unavoidable reaction in the current kinetic system. Based on our experimental observations, we suggest that $CH_2OO$ should react quickly with IO. Fortunately, the amount of IO is controlled by the initial amount of I atoms under our experimental conditions where $[O_3] \gg [I]$. Thus, the kinetics of the $CH_2OO$ decay is still pseudo-1st-order and the observed $k_{eff}$ is linear to $[O_3]$ for $[O_3] > 0$.

The measured rate coefficient of $CH_2OO$ reaction with $O_3$ is $(6.72\pm0.46)\times10^{-14}$ $cm^3\,sec^{-1}$. This value differs from the predicted values in the literature by one order of magnitude or more, suggesting the need of multireference treatment for the quantum chemistry calculations. While theory has made good predictions or estimations for a number of reaction rates of Criegee intermediates with water, $H_2S$ and some other molecules,[24] it seems still tricky to calculate the rate of $CH_2OO$ reaction with $O_3$. We hope the measured results of the present work could benefit future theoretical calculations for this reaction.

This measurement also indicates that the reaction of $CH_2OO$ with $O_3$ should play a role in laboratory studies of ozonolysis, where the early-time decay of $CH_2OO$ may be controlled by its reactions with $O_3$ and with the used alkene molecules. The rate coefficients of $CH_2OO$ reactions with simple alkenes have been reported by Buras et al.[14] to be $2\times10^{-15}$ to $11\times10^{-15}$ $cm^3\,sec^{-1}$, which are smaller than that for the $O_3$ reaction. As a result, if one wish to produce a higher steady-state concentration of $CH_2OO$ in ozonolysis experiments, an ozone concentration lower than that of alkene may be desired to slow down the decay of $CH_2OO$ due to its reaction with $O_3$. For an ozonolysis study with 1 ppm of $O_3$, the effective decay rate of $CH_2OO$ by $O_3$ is about 2 $s^{-1}$, which is slightly larger than the thermal decomposition rate of $CH_2OO$ (0.2 $s^{-1}$).[42] On the other hand, the reaction of $CH_2OO$ with atmospheric ozone is relatively slower ($\sim 0.17$ $s^{-1}$ for 100 ppbv $O_3$) than other $CH_2OO$ sinks, such as the reaction with water dimer ($> 1000$ $s^{-1}$),[17] in the atmosphere.

## ASSOCIATED CONTENT

The following files are available free of charge.

Supporting information (file type: PDF): experimental conditions, kinetic modeling, supplementary figures and tables.

## AUTHOR INFORMATION

Corresponding Author

E-mail: ypchang@mail.nsysu.edu.tw

Notes

The authors declare no competing financial interest.

## ACKNOWLEDGMENTS

This work is supported by the Ministry of Science and Technology, Taiwan (MOST103-2113-M-001-019-MY3) and Academia Sinica. We thank Mr. Wen Chao for helping the measurements. We also thank Dr. Mica Smith and Dr. Kaito Takahashi for discussions about the reaction of $CH_2OO$ with IO.

## REFERENCES

1  R. Criegee, Angew. Chem. Int. Ed. Engl., 1975, **14**, 745.
2  C.A. Taatjes, D.E. Shallcross, and C.J. Percival, Phys. Chem. Chem. Phys., 2014, **16**, 1704.
3  D.L. Osborn and C.A. Taatjes, Int. Rev. Phys. Chem., 2015, **34**, 309.
4  Y.-P. Lee, J. Chem. Phys., 2015, **143**, 020901.
5  O. Welz, J.D. Savee, D.L. Osborn, S.S. Vasu, C.J. Percival, D.E. Shallcross, and C.A. Taatjes, Science, 2012, **335**, 204.
6  N.M. Donahue, G.T. Drozd, S.A. Epstein, A.A. Presto, and J.H. Kroll, Phys. Chem. Chem. Phys., 2011, **13**, 10848.
7  A. Novelli, L. Vereecken, J. Lelieveld, and H. Harder, Phys Chem Chem Phys , 2014,**16**, 19941.
8  F. Liu, J.M. Beames, A.S. Petit, A.B. McCoy, and M.I. Lester, Science, 2014, **345**, 1596.
9  N.M. Kidwell, H. Li, X. Wang, J.M. Bowman, and M.I. Lester, Nat. Chem., 2016, **8**, 509.
10 C.A. Taatjes, O. Welz, A.J. Eskola, J.D. Savee, A.M. Scheer, D.E. Shallcross, B. Rotavera, E.P.F. Lee, J.M. Dyke, D.K.W. Mok, D.L. Osborn, and C.J. Percival, Science, 2013, **340**, 177.
11 L. Sheps, A.M. Scully, and K. Au, Phys Chem Chem Phys, 2014, **16**, 26701.

12 W.-L. Ting, C.-H. Chang, Y.-F. Lee, H. Matsui, Y.-P. Lee, and J.J.-M. Lin, J. Chem. Phys., 2014, **141**, 104308.

13 Y.-T. Su, H.-Y. Lin, R. Putikam, H. Matsui, M.C. Lin, and Y.-P. Lee, Nat. Chem., 2014, **6**, 477.

14 Z.J. Buras, R.M.I. Elsamra, A. Jalan, J.E. Middaugh, and W.H. Green, J. Phys. Chem. A, 2014, **118**, 1997.

15 D. Stone, M. Blitz, L. Daubney, N.U.M. Howes, and P. Seakins, Phys Chem Chem Phys, 2014, **16**, 1139.

16 O. Welz, A.J. Eskola, L. Sheps, B. Rotavera, J.D. Savee, A.M. Scheer, D.L. Osborn, D. Lowe, A. Murray Booth, P. Xiao, M. Anwar H. Khan, C.J. Percival, D.E. Shallcross, and C.A. Taatjes, Angew. Chem. Int. Ed., 2014, **53**, 4547.

17 W. Chao, J.-T. Hsieh, C.-H. Chang, and J.J.-M. Lin, Science, 2015, **347**, 751.

18 H.-L. Huang, W. Chao, and J.J.-M. Lin, Proc. Natl. Acad. Sci., 2015, **112**, 10857.

19 M.C. Smith, C.-H. Chang, W. Chao, L.-C. Lin, K. Takahashi, K.A. Boering, and J.J.-M. Lin, J. Phys. Chem. Lett., 2015, **6**, 2708.

20 R. Chhantyal-Pun, A. Davey, D.E. Shallcross, C.J. Percival, and A.J. Orr-Ewing, Phys Chem Chem Phys, 2015, **17**, 3617.

21 E.S. Foreman, K.M. Kapnas, and C. Murray, Angew. Chem., 2016, **128**, 10575.

22 L.-C. Lin, H.-T. Chang, C.-H. Chang, W. Chao, M.C. Smith, C.-H. Chang, J. Jr-Min Lin, and K. Takahashi, Phys Chem Chem Phys, 2016, **18**, 4557.

23 M.C. Smith, W. Chao, M. Kumar, J.S. Francisco, K. Takahashi, and J.J.-M. Lin, J. Phys. Chem. A, 2017, **121**, 938.

24 J. Jr-Min Lin and W. Chao, Chem Soc Rev, 2017. DOI: 10.1039/c7cs00336f

25 L. Sheps, B. Rotavera, A.J. Eskola, D.L. Osborn, C.A. Taatjes, K. Au, D.E. Shallcross, M.A.H. Khan, and C.J. Percival, Phys Chem Chem Phys, 2017, **19**, 21970.

26 Z.C.J. Decker, K. Au, L. Vereecken, and L. Sheps, Phys Chem Chem Phys, 2017, **19**, 8541.

27 Y. Liu, F. Liu, S. Liu, D. Dai, W. Dong, and X. Yang, Phys Chem Chem Phys, 2017, **19**, 20786.

28 C.A. Taatjes, Annu. Rev. Phys. Chem., 2017, **68**, 183.

29 R.L. Caravan, M.A.H. Khan, B. Rotavera, E. Papajak, I.O. Antonov, M.-W. Chen, K. Au, W. Chao, D.L. Osborn, J.J.-M. Lin, C.J. Percival, D.E. Shallcross, and C.A. Taatjes, Faraday Discuss, 2017, **200**, 313.

30 R. Chhantyal-Pun, M.R. McGillen, J.M. Beames, M.A.H. Khan, C.J. Percival, D.E. Shallcross, and A.J. Orr-Ewing, Angew. Chem. - Int. Ed., 2017, **56**, 9044.

31 R. Chhantyal-Pun, O. Welz, J.D. Savee, A.J. Eskola, E.P.F. Lee, L. Blacker, H.R. Hill, M. Ashcroft, M.A.H. Khan, G.C. Lloyd-Jones, L. Evans, B. Rotavera, H. Huang, D.L. Osborn, D.K.W. Mok, J.M. Dyke, D.E. Shallcross, C.J. Percival, A.J. Orr-Ewing, and C.A. Taatjes, J. Phys. Chem. A, 2017, **121**, 4.

32 J.H. Kroll and J.H. Seinfeld, Atmos. Environ., 2008, **42**, 3593.

33 H.G. Kjaergaard, T. Kurtén, L.B. Nielsen, S. Jørgensen, and P.O. Wennberg, J. Phys. Chem. Lett., 2013, **4**, 2525.

34 W. Wei, R. Zheng, Y. Pan, Y. Wu, F. Yang, and S. Hong, J. Phys. Chem. A, 2014, **118**, 1644.

35 L. Vereecken, H. Harder, and A. Novelli, Phys. Chem. Chem. Phys., 2014, **16**, 4039.

36 L. Vereecken, A.R. Rickard, M.J. Newland, and W.J. Bloss, Phys Chem Chem Phys, 2015, **17**, 23847.

37 Y.-P. Chang, A.J. Merer, H.-H. Chang, L.-J. Jhang, W. Chao, and J.J.-M. Lin, J. Chem. Phys., 2017, **146**, 244302.

38 B. Jin, M.-N. Su, and J.J.-M. Lin, J. Phys. Chem. A, 2012, **116**, 12082.

39 W.-L. Ting, Y.-H. Chen, W. Chao, M.C. Smith, and J.J.-M. Lin, Phys Chem Chem Phys, 2014, **16**, 10438.

40 J.H. Kroll, J.S. Clarke, N.M. Donahue, and J.G. Anderson, J. Phys. Chem. A, 2001, **105**, 1554.

41 M.E. Tucceri, T.J. Dillon, and J.N. Crowley, Phys Chem Chem Phys, 2005, **7**, 1657.

42 T. Berndt, R. Kaethner, J. Voigtländer, F. Stratmann, M. Pfeifle, P. Reichle, M. Sipilä, M. Kulmala, and M. Olzmann, Phys. Chem. Chem. Phys., 2015, **17**, 19862-19873.